**Remote, bivariate expert elicitation to determine the prior probability distribution for sample size calculation in a Bayesian non-inferiority multicenter randomized controlled trial: A single dose of dexamethasone at 0.60 mg/kg of body weight versus 0.15 mg/kg for the treatment of croup (Croup Dosing Trial)**

**Author List**: Arlene Jiang, Alex Aregbesola, Apoorva Gangwani, Terry P. Klassen, Amy C. Plint, Elisabete Doyle, William Craig, Mohamed Eltorki, Banke Oketola, Hoda Badra, Yongdong Ouyang*, Anna Heath*

* Contributed equally

Affiliations:

Arlene Jiang (arlene.jiang@sickkids.ca): Child Health Evaluative Sciences, The Hospital for Sick Children, Toronto, Ontario, Canada

Alex Aregbesola (alex.aregbesola@umanitoba.ca): Children's Hospital Research Institute of Manitoba, Winnipeg, Manitoba, Canada; Department of Pediatrics and Child Health, Max Rady College of Medicine, Rady Faculty of Health Sciences, University of Manitoba, Winnipeg, Canada

Apoorva Gangwani (agangwani@chrim.ca): Children's Hospital Research Institute of Manitoba, Winnipeg, Manitoba, Canada

Terry P. Klassen (terry.klassen@usask.ca): College of Medicine, University of Saskatchewan and Jim Pattison Children's Hospital, Saskatchewan Health Authority, Saskatoon, Saskatchewan, Canada

Amy C. Plint (plint@cheo.on.ca): Department of Pediatrics, University of Ottawa, Ottawa, Ontario, Canada; Division of Emergency Medicine, Children's Hospital of Eastern Ontario, Ottawa, Ontario, Canada

Elisabete Doyle (edoyle@exchange.hsc.mb.ca): Department of Pediatrics and Child Health, University of Manitoba, Winnipeg, Manitoba, Canada

William Craig (wcraig@ualberta.ca): Department of Medicine and Dentistry, University of Alberta, Edmonton, Alberta, Canada

Mohamed Eltorki (eltorkim@mcmaster.ca): Cumming School of Medicine, Department of Pediatrics, University of Calgary, Calgary, Alberta, Canada

Banke Oketola (boketola@chrim.ca): Children's Hospital Research Institute of Manitoba, Winnipeg, Manitoba, Canada; University of Manitoba, Winnipeg, Manitoba, Canada

Hoda Badran (hoda.badran@umanitoba.ca): Children's Hospital Research Institute of Manitoba, Winnipeg, Manitoba, Canada; Department of Pediatrics and Child Health, Max Rady College of Medicine, Rady Faculty of Health Sciences, University of Manitoba, Winnipeg, Canada

Yongdong Ouyang (yongdong.ouyang@RoswellPark.org): Department of Biostatistics and Bioinformatics, Roswell Park Comprehensive Cancer Center, Buffalo, New York, United States

Anna Heath (anna.heath@sickkids.ca): Child Health Evaluative Sciences, The Hospital for Sick Children, Toronto, Toronto, Ontario, Canada; Division of Biostatistics, Dalla Lana School of Public Health, University of Toronto, Toronto, Ontario, Canada; Department of Statistical Science, University College London, London, United Kingdom

**Corresponding Author:**

Arlene Jiang

The Hospital for Sick Children, Toronto

686 Bay Street, Toronto ON M5G 0A4

Email: arlene.jiang@sickkids.ca

**Abstract**

**Background:** Prior distributions must be specified for the parameters of interest in a Bayesian clinical trial. When existing evidence on the effects of the trial interventions is limited or inconclusive, prior distributions can be constructed with expert elicitation. However, conventional elicitation requires face-to-face interactions and intensive pre-elicitation training, which can be infeasible and costly. Our remote elicitation was based on an established expert elicitation methodology, and we incorporated bivariate prior distributions to introduce dependencies between the elicited probabilities. We aimed to elicit a prior distribution for the Croup Dosing Trial, which assesses the efficacy of two separate doses of dexamethasone on the number of return visits to the emergency department within 7 days in children with croup. This trial evaluates the non-inferiority of 0.15 mg/kg of dexamethasone, compared to the standard dose of 0.60 mg/kg to treat croup.

**Methods:** We conducted three remote workshops to elicit expert beliefs on the efficacy of the two doses of dexamethasone. Each workshop consisted of two survey rounds, separated by a group discussion. Prior to the workshop, experts reviewed the same current literature that was provided on the effects of the two doses of dexamethasone. Beliefs were aggregated using expert-specific bivariate distributions with latent effects. The aggregated distribution, along with the surveyed non-inferiority margin, determined the sample size for the Bayesian non-inferiority trial design.

**Results:** Twelve emergency medicine physicians participated in our remote elicitation exercise. The elicitation generated a prior distribution centered at 6% for the 0.60 mg/kg dose and 8% for the 0.15 mg/kg dose. The aggregated prior distribution produced a sample size of 1850, based on a non-inferiority margin of 4%.

**Conclusions:** We elicited a prior distribution that incorporated past evidence and expert opinion. The elicited prior is consistent with previous literature on the efficacy of the dexamethasone doses in treating croup. Our approach demonstrates the feasibility of remotely eliciting bivariate distributions to design clinical trials.

**Trial Registration:** NCT06272383 (Registered May 8, 2024)

**Keywords:** Expert Elicitation, Bayesian Statistics, Randomized Controlled Trials, Trial Design, Non-Inferiority Trial, Sample Size Determination, Prior Probability Distribution

## Background

Clinical trials are increasingly designed and evaluated using Bayesian methods due to their interpretability and increased efficiency.[1] Bayesian methods require the specification of prior distributions for the parameters of interest that reflect the uncertainty of decision makers.[2] Existing evidence from previous trials is often incorporated into prior distributions, which are then combined with study data to generate trial conclusions. However, this approach can be limited when similar trials have not been previously conducted.

Alternatively, a prior distribution based on expert opinion can be formally obtained through expert elicitation.[3] Elicitation is the scientific process of converting content experts' opinions into statistical distributions to quantify the uncertainty on the parameters of interest.[4] Elicitation is often used to support decision-making in areas with limited or inconclusive evidence.[5] In the context of clinical trials, elicitation allows stakeholders to directly incorporate their uncertainty about the effects of the study interventions while planning the trial.[6] Over the past few decades, a growing number of frameworks and applications have been developed for the process of elicitation.[5,7–11]

Croup is a common respiratory disease in children, associated with frequent emergency department (ED) visits.[12] Usual symptoms of croup include a seal-like barking cough and a whistling noise while breathing.[12] Past studies have demonstrated that glucocorticoids are effective in managing symptoms and reducing healthcare costs (i.e. hospitalization, admission to intensive care unit, return visits).[13–15] However, the course of glucocorticoids prescribed for croup is associated with adverse events such as gastrointestinal bleeding, pneumonia, and sepsis.[16] The Croup Dosing Trial will assess the efficacy of a lower dose of glucocorticoids for the treatment of croup in children. The trial will test the non-inferiority of a lower dose of dexamethasone compared to the standard dose on the number of revisits to the ED within 7 days. This trial aims to provide definitive evidence for how croup should be treated in children.

The Croup Dosing Trial will take place in Canada, USA, New Zealand, and Australia. Therefore, a remote elicitation involving experts from different countries would generate a more representative prior for the trial population. The goal of the elicitation is to produce a prior distribution that would be used for Bayesian sample size determination and the analysis of the trial data. This study describes the design and conduct of the elicitation to derive a prior distribution for the Croup Dosing Trial. Traditional elicitation requires intensive pre-elicitation training and face-to-face sessions.[17] This approach may be infeasible given the time, geographical, and cost restrictions in assembling a group of experts. Lan et al. successfully implemented a remote expert elicitation for a clinical trial in bronchiolitis in infants termed the BIPED trial,[18] which addressed these limitations by holding multiple remote elicitation workshops. Conducting multiple workshops accommodated different time zones and schedules, which encouraged experts from Canada, the USA, Australia, and New Zealand to participate. This added to the heterogeneity of expert opinions, which is practically and computationally advantageous.[19] This approach is based on the validated IDEA (Investigate Discuss Estimate Aggregate) protocol,[5] which we plan to leverage in the Croup Dosing Trial expert elicitation. The BIPED study assumed independence between the probability of efficacy in each treatment arm. However, experts likely used their estimate of one treatment arm to inform the other, also known as anchoring.[20,21] One approach that avoids anchoring is to elicit two distinct quantities, such as the response in the control arm and the treatment effect.[22,23] These quantities are considered independent, and thus these estimates are aggregated separately. However, a treatment effect such as an odds ratio is not directly observable and is therefore more difficult to elicit than the response in either study arm.[7] Because of this, we elicited the probability of the outcome in our study arms, addressing the dependence between these quantities with a bivariate prior distribution for each expert. Thall et al. implemented individual expert latent effects to introduce a correlation between the expert-specific prior distributions in each treatment arm.[24] This approach was used to produce a prior distribution for a trial studying pediatric idiopathic nephrotic syndrome. We planned to incorporate this bivariate approach to account for possible dependencies between treatment arm estimates and conduct remote expert elicitation for the Croup Dosing Trial.

**Methods**

<u>Croup Dosing Trial</u>

The Croup Dosing Trial is a phase III, double-blind, multicenter, randomized controlled non-inferiority trial comparing two treatment regimens for children presenting to the ED with croup within the Pediatric Emergency Research Canada (PERC). Children between 6 months to 5 years with a clinical diagnosis of croup will be eligible. The trial will determine whether a 0.15 mg/kg dose of dexamethasone is non-inferior to the standard dose of 0.60 mg/kg. The primary outcome of this trial is the number of return visits within 7 days of initial presentation to the ED. Patients will be allocated in a ratio of 1:1 to receive either dose of dexamethasone.

Research Ethics Approval

The vanguard and elicitation components of the Croup Dosing Trial have received ethics approval of Health Canada, University of Manitoba Health Research Ethics Board, and Shared Health on behalf of the Children's Hospital of Winnipeg. Use of the elicitation data was approved by the ethics committee at the Hospital for Sick Children.

Designing the Elicitation

***Key Parameters***

In our elicitation, the key parameters were the probability of return visits to the ED within 7 days after initial presentation, following treatment with either 0.60 mg/kg or 0.15 mg/kg of dexamethasone. We denote these two probabilities as *p1* and *p2*, respectively. For each expert, we assume that these probabilities follow beta distributions, a common distributional choice for modeling probabilities.[25]

Key parameters were elicited using the workshop methodology described in Lan et al.[18] We provided experts with a case study of a child with croup who met inclusion/exclusion criteria for the Croup Dosing Trial (Supplementary File 1). The case study describes a 1-year-old boy with symptoms indicative of a mild to moderate croup diagnosis. Experts were asked to determine the number of patients, out of 100, with characteristics like this patient who would revisit the ED for the two doses of dexamethasone.

In addition to the elicited probability estimates, we considered the effect of each expert's professional experience on their judgement of *p1* and *p2.* Therefore, we surveyed the professional experience of the experts (Supplementary File 2) and incorporated this information into aggregating expert opinions for the final prior distribution.[24]

### *Online Elicitation Tool*

To support our elicitation, we created an online, interactive tool using R and the shiny package (https://apoorva-gangwani.shinyapps.io/apoorva_elicitation/).[26,27] This tool allowed experts to visualize and send in their elicitation estimates. Experts were prompted to provide the lower and upper plausible values, followed by their most plausible value that subjectively described their uncertainty about the probability of ED revisits within 7 days of initial presentation to the ED at both doses of dexamethasone. The lower and upper plausible values were assumed to represent the 95% central credible interval of a beta distribution, while the most plausible value was assumed to represent the mode. These values were fit to beta distributions using betaExpert.[28] After selecting elicitation values, our tool provided experts with a summary plot of the corresponding beta distribution. This real-time visualization allowed experts to adjust the fitted beta distribution until they deemed their beliefs to be sufficiently elicited. Finally, experts were prompted to send their elicitation values on a submission page hosted using REDCap.[29,30]

### *Selecting the Experts*

We identified experts from Canada, USA, Australia, and New Zealand, to determine a prior representative of the regions where the Croup Dosing Trial will be conducted. Experts were primarily recruited from PERC, a national network, and the Pediatric Emergency Research Network (PERN),[31] an international organization. Participants could be selected if they (i) were identified as experts in croup and its treatment and (ii) had experience in pediatric emergency medicine. Participants were ineligible if they had (i) prior involvement with the Croup Dosing

Trial or (ii) authored any studies that preferred dexamethasone at 0.60 mg/kg or 0.15 mg/kg. We aimed to recruit 10 to 15 experts for a diversity in geographical representation and experience.[5,32]

### *Aggregating the Prior Distribution*

To derive the final prior distribution, we aggregated individual distributions with methods developed by Thall et al.[24] These methods construct a bivariate prior for each expert, introducing a correlation between responses for the two doses. We first derived marginal distributions for each expert's estimate for *p1* and *p2*. Covariates related to physician's expertise were added during the fitting process to explain the variability between experts' responses. An expert-specific correlation between two marginal distributions was constructed using latent expert effects. For this regression, we used noninformative priors for the covariate effects ($N(0, 10)$), the intercepts ($N(0,100)$), the standard deviation of the latent effects ($IG(1,1)$), and the correlation between the latent effects ($Unif(-1,1)$). Each physician's bivariate prior was weighted evenly in the final aggregate.

<u>The Elicitation</u>

### *Pre-Workshop Materials*

Before the workshop, experts were sent a set of readings to prepare for the workshop.[33,34] These readings included one clinical trial and one meta-analysis, presenting past and latest evidence on the efficacy of 0.60 mg/kg and 0.15 mg/kg of dexamethasone in the treatment of croup in children. These studies were sent to complement the experts' existing knowledge on this topic.

### *Remote, real-time expert elicitation workshop*

We conducted a 90-minute workshop, standardized with a script (Supplementary File 3),[35] over Zoom in April 2024. The same workshop was held a total of three times to accommodate different time zones and the schedules of working physicians. Three to four statistical or medical experts facilitated each workshop (AJ, AG, AA, AH). The workshop began with an introduction to the context of the trial and Bayesian statistics. We then demonstrated the elicitation tool with an example dummy question (Supplementary File 3).

The workshop format was adapted from the IDEA protocol.[5] The IDEA protocol outlines four stages to an elicitation: (1) participants individually form estimates on the quantities of interest, (2) participants discuss their reasoning, (3) participants individually re-estimate the quantities of interest, and (4) individual estimates are aggregated. Accordingly, each of our elicitation workshops included two rounds of questions, separated by a group discussion. During each round, experts individually submitted their prior distributions on the probability of revisiting the ED in each treatment arm. During the group discussion, experts were presented with deidentified boxplots containing their answers from round 1 (Supplementary File 4). Experts were given 15 minutes to discuss why their answers were different. After reflecting on the discussion at workshop 1, the facilitating team prepared a set of questions to guide the discussion at the other two workshops (Supplementary File 3). The group discussion provided experts with an opportunity to reconsider their initial beliefs.

### Post-workshop

Post-workshop, experts were sent their individual, workshop-aggregate, and aggregated distributions for the probability of ED revisits in each treatment arm. Experts were invited to comment on these findings over email.

<u>Bayesian Sample Size Determination</u>

### Non-Inferiority Margin

The non-inferiority margin was estimated by surveying PERC physicians at the 11 prospective sites for the Croup Dosing Trial in December 2025 (Supplementary File 5). The median of the survey answers was taken as the margin.

### Sample Size Determination

Our sample size was determined using assurance, which required simulation in R using *INLA*.[26,36] We generated binomial data, representing the number of return visits in each arm, with probabilities given by each pair of samples in the aggregated *p1* and *p2* distributions. These simulated data were fit to general linear models with a prior distribution for the difference between *p2* and *p1*. We declared non-inferiority when 95% of the posterior mean differences were smaller than the non-inferiority margin. To evaluate the null scenario (analogous to type I error), we repeated this process with binomial data generated with probabilities from *p1* and with probabilities from *p1* plus the non-inferiority margin. We chose a sample size that provided 80% of the maximum possible assurance, i.e. relative assurance, and 5% assurance in the null scenario. Simulations were repeated 1000 times.

## Results

Elicitation Workshop

### Baseline Characteristics

We invited a total of 33 experts from Australia (n=5), Canada (n=17), New Zealand (n=4) and the USA (n=7). Experts declined participation because they were on vacation, had scheduling conflicts, or had large time differences. We recruited 12 clinicians from Canada (n=8) and the USA (n=4) to participate in our workshops. Table 1 displays the characteristics of these experts from our survey of their professional experience. Experts had 11 to 38 years of experience in practicing medicine. In the past year, all physicians had prescribed 0.60 mg/kg of dexamethasone, while only two had prescribed 0.15 mg/kg of dexamethasone. Most clinicians had received training in clinical trial methodology and statistics.

Table 1. Baseline characteristics of the expert elicitation participants.

| Country of practice | Canada | 8 |
|---|---|---|
| | USA | 4 |
| How many years have you practiced medicine? | 11-15 | 3 |
| | 16-20 | 2 |
| | 21-25 | 3 |
| | 26-30 | 1 |
| | 31-35 | 2 |
| | 36-40 | 1 |
| Have you prescribed 0.60 mg/kg of dexamethasone for croup over the last year? | Yes | 12 |
| | No | 0 |
| Have you prescribed 0.15 mg/kg of dexamethasone for croup over the last year? | Yes | 2 |
| | No | 10 |
| What's the maximum dose (mg) of dexamethasone that you have prescribed for croup in the last year? | 6 | 1 |
| | 10 | 5 |
| | 12 | 2 |
| | 15 | 1 |
| | 16 | 3 |
| Have you received any training in clinical trial methodology? | Yes | 10 |
| | No | 2 |
| Have you received any training in statistics? | Yes | 11 |
| | No | 1 |

Prior Distributions

Figure 1 and Table 2 show the individual's prior beliefs on the efficacy of each dose of dexamethasone before and after discussion. Individual responses varied and tended to decrease in variability from round 1 to round 2. Furthermore, many experts modified their best estimate from round to round. In general, experts believed that there would be a higher probability of revisiting the ED in children treated with 0.15 mg/kg of dexamethasone. Figure 2 shows the aggregated prior distributions from all experts for each round of questions. Similar to the previous figure, these graphs indicate the decrease in variability of the answers from round 1 to round 2. The aggregated distributions from round 2 were taken to be the final prior distributions used for sample size determination (Table 3). Figure 3 presents the risk difference between the probability of revisits under 0.15 mg/kg versus 0.60 mg/kg of dexamethasone from the joint distribution. The narrow central tendency demonstrates the correlation between *p1* and *p2*. The posterior estimates have a correlation of 0.40, indicating a moderate positive association between the two elicited quantities. Experts did not provide additional comments over email.

Figure 1. Individual prior distributions for the probability of ED revisit under 0.60 mg/kg and 0.15 mg/kg of dexamethasone by round.

Table 2. Mean and standard deviations (SD) of individual prior distributions for the probability of ED revisit under 0.60 mg/kg and 0.15 mg/kg of dexamethasone.

| | Round 1 0.60mg/kg | | Round 2 0.60mg/kg | | Round 1 0.15mg/kg | | Round 2 0.15mg/kg | |
|---|---|---|---|---|---|---|---|---|
| Expert | Mean | SD | Mean | SD | Mean | SD | Mean | SD |
| 1 | 9.43% | 6.33% | 11.37% | 7.61% | 10.79% | 7.05% | 12.74% | 9.34% |
| 2 | 6.16% | 3.65% | 7.53% | 4.36% | 7.90% | 3.92% | 8.63% | 5.66% |
| 3 | 2.47% | 1.08% | 3.61% | 1.49% | 2.47% | 1.08% | 3.61% | 1.73% |
| 4 | 21.75% | 6.94% | 14.58% | 5.10% | 26.64% | 7.88% | 16.68% | 4.57% |
| 5 | 5.91% | 3.36% | 5.81% | 2.21% | 8.32% | 5.21% | 9.12% | 5.24% |
| 6 | 9.87% | 4.40% | 8.74% | 3.98% | 13.71% | 5.10% | 11.67% | 4.39% |
| 7 | 14.36% | 7.98% | 10.88% | 7.71% | 22.74% | 8.77% | 15.80% | 13.03% |
| 8 | 11.24% | 3.91% | 8.01% | 2.92% | 16.92% | 6.13% | 11.24% | 2.84% |
| 9 | 6.85% | 2.47% | 6.53% | 1.91% | 11.43% | 4.21% | 11.43% | 4.21% |
| 10 | 7.51% | 5.05% | 5.38% | 4.16% | 10.88% | 7.71% | 6.16% | 3.03% |
| 11 | 6.78% | 3.50% | 6.78% | 3.50% | 9.26% | 4.64% | 7.45% | 2.62% |
| 12 | 8.05% | 4.93% | 6.78% | 3.50% | 10.16% | 4.78% | 8.97% | 2.03% |

Figure 2. Aggregated prior distributions from individual expert's elicited beta prior distribution for the probability of ED revisit by round.

Table 3. Aggregated prior distributions from individual expert's elicited beta prior distribution for the probability of ED revisit from round 2.

|            | Mean   | Median | SD     |
|------------|--------|--------|--------|
| *p1*       | 8.01%  | 6.44%  | 6.22%  |
| *p2*       | 1.00%  | 8.19%  | 7.27%  |
| *p2 – p1*  | 1.98%  | 1.52%  | 7.42%  |

Figure 3. Risk difference between ED revisits for 0.15 mg/kg vs. 0.60 mg/kg of dexamethasone.

Bayesian Sample Size Determination

Our survey of 60 physicians found a non-inferiority margin of 4%. To determine the trial's sample size, binomial data was generated with probabilities of 6.44% (*p1*) and 8.19% (*p2*) (Table 3). We approximated the prior distribution for the difference between *p2* and *p1* with a Pearson IV distribution (m=1.91, nu=-0.36, location=0.01, scale=0.08) to best capture the shape of the distribution presented in Figure 3 Table 3). A sample size of 1850 provided 80% relative assurance and 5% assurance in the null scenario.

**Discussion**

In this study, we implemented a remote expert elicitation exercise to develop a bivariate prior distribution for the Croup Dosing Trial. This prior distribution was used to determine the sample size and will be used to analyze the trial. Our elicited prior had median ED revisit probabilities of 6.4% for the 0.60 mg/kg dose and 8.2% for the 0.15 mg/kg dose. These probabilities are consistent with Parker and Cooper,[34] who found that 5.9% and 8.8% of patients revisited the ED when treated with 0.60 mg/kg and 0.15 mg/kg of dexamethasone, respectively. Our elicited probabilities are not directly comparable to the meta-analysis,[33] which accounted for return visits to both the ED and family doctors. Consistent with the meta-analysis, however, which reported return visit rates of 19% and 21% for 0.60 mg/kg and 0.15 mg/kg, we found a comparable risk difference between treatment arms. The similarities between our elicited probabilities and past evidence demonstrate the reliability of our prior distribution.

During our elicitation, we found a shift in expert beliefs from round to round. Between survey rounds, the group discussions provided experts with an opportunity to comment on how they chose their estimates. Experts re-evaluated their interpretation of the lower and upper plausible values, as well as how the primary outcome relates to the clinical progression of croup. Physicians commented on factors that would impact the efficacy of the doses of dexamethasone such as time of year and virulence of the virus. Another consideration was how regional differences in access to primary care would affect the number of children who would visit their family doctor rather than return to the ED. Experts subsequently provided less varied answers during the second survey round. The topics discussed were consistent with the goals of the discussion stage from the IDEA protocol of encouraging critical thinking, discussing evidence, and clarifying linguistic ambiguity. A change in expert beliefs from each round was also reported in the BIPED study.[18] This demonstrates the value of the real-time discussion component to our elicitation.

Prior to the elicitation, there was existing evidence on the efficacy of the Croup Dosing Trial interventions. As noted by one of our experts, the meta-analysis could have been used as the prior distribution for the trial.[33] While this meta-analysis presents the best evidence on the efficacy of the two doses of dexamethasone on return visits, there are only three studies in this area.[37,38,34] This relatively small number of studies may result in imprecise estimates.[39,40] Therefore, we opted to use expert opinion to develop priors. As demonstrated during the group discussion, experts have a broader understanding of the context of their practice and how it differs from other settings. This may allow experts to produce estimates that generalize better than past single-center trials.[34,37,38] By consulting with multiple physicians, we also included perspectives from heterogeneous settings differing in case mix.[19] Furthermore, our elicitation exercise asked physicians to consider past empirical evidence, which effectively combined data with expert knowledge. Accordingly, five of our experts brought up using past data and their own experience to inform their estimates during the group discussion. Taken together, our expert elicitation allowed us to generate a more robust prior that accounted for both existing data and physicians' expertise.

The duration of the workshops is a strength and limitation of our methodology. Our remote 90-minute workshops could accommodate the schedules of multiple physicians. Conversely, the length of these workshops limited how much training we provided to our experts. For instance, we did not review potential biases or subjective probability, as recommended by many frameworks.[7] Future elicitations should aim to include such training to derive high-quality prior distributions. Another limitation to our elicitation exercise is that it is unclear why experts changed their opinions following the group discussion. While group discussions support the calibration of expert judgement, they may also introduce biases. For instance, an expert may uncritically adopt the opinions of others, also known as groupthink.[41] This would result in a prior distribution that does not truly reflect the collective experts' opinion, but rather overrepresents certain group members' opinions. To reduce the effects of groupthink, future studies should document expert rationale during the second round of elicitation to increase transparency into each expert's thought process.

Another limitation is that we did not evaluate experts' experience with our elicitation. Future studies should collect whether experts feel adequately trained and what experts think about the elicitation exercise to identify areas that could be improved. Overall, this study successfully demonstrates the remote elicitation of a bivariate prior for the Croup Dosing Trial. The elicited prior is consistent with previous literature on the efficacy of the dexamethasone doses in treating croup. This prior distribution combined empirical evidence with expert opinion and determined the sample size of the Croup Dosing Trial. Our study supports the use of prior elicitation for designing future clinical trials.

## List of Abbreviations

IDEA: Investigate Discuss Estimate Aggregate; ED: emergency department; PERN: Pediatric Emergency Research Network; PERC: Pediatric Emergency Research Canada; SD: standard deviation

## Declarations

**Ethics approval and consent to participate:**

The elicitation of the Croup Dosing Trial has ethics approval from Health Canada (# CDS0923) as well as University of Manitoba Health Research Ethics Board & Shared Health (# HS26166 (B2023:092)) approval. The Hospital for Sick Children Research Ethics Board approved a secondary use of the elicitation data (# 1000081720). Implied consent was used in this study.

**Consent for publication:** Not applicable.

**Availability of data and materials:** All elicited data generated during this study are included in this published article. Individual survey data are not publicly available due to the ease of identifying individuals but are available from the corresponding author on reasonable request.

**Competing interests:** The authors declare that they have no competing interests.

**Funding:** AA is supported by a Research Manitoba Operating Grant. AJs time was also supported by the Canadian Network of Statistical Training in Trials. TPK is funded by a Canada Research Chair in Clinical Trials and AH is funded by a Canada Research Chair in Statistical Trial Design.

**Authors' contributions:**

AH designed the elicitation exercise with the support of TPK. AG created the elicitation tool. ACP, ED, WC, ME, TPK helped develop the elicitation case scenario. AG and AA managed expert recruitment. AJ, AA, AG, and AH facilitated the elicitation. AG collected data and managed communication with the experts. BO and HB participated in the elicitation workshops. AJ analyzed the data and drafted the manuscript. AA, AG, and YO contributed substantive revisions. YO supervised the writing process. All authors approved the final manuscript.

**Acknowledgements**: We would like to thank all elicitation workshop participants, James Chamberlain, Todd A. Florin, Jocelyn Gravel, Lise Nigrovic, Martin Osmond, Tania Principi, Vikram Sabhaney, Graham Thompson, Jennifer Thull-Freedman, Leah Tzimenatos, Rob Woods, Bruce Wright. The authors also like to thank the administrative staff of the Children's Hospital Research Institute of Manitoba for their support of our research work and appreciate the financial contributions from the Children's Hospital Foundation of Manitoba toward REACH (Research Into the Enhancement of Acute care for Children' Health).

**References**

1.  Berry DA. Bayesian clinical trials. Nat Rev Drug Discov. 2006 Jan;5(1):27–36.

2.  Ruberg SJ, Beckers F, Hemmings R, Honig P, Irony T, LaVange L, et al. Application of Bayesian approaches in drug development: starting a virtuous cycle. Nat Rev Drug Discov. 2023 Mar;22(3):235–50.

3.  van de Schoot R, Depaoli S, King R, Kramer B, Märtens K, Tadesse MG, et al. Bayesian statistics and modelling. Nat Rev Methods Primers. 2021 Jan 14;1(1):1–26.

4.  Dias LC, Morton A, Quigley J, editors. Elicitation: The Science and Art of Structuring Judgement [Internet]. 1st ed. Springer Cham; 2018 [cited 2024 June 20]. 542 p. Available from: https://doi.org/10.1007/978-3-319-65052-4

5.  Hemming V, Burgman MA, Hanea AM, McBride MF, Wintle BC. A practical guide to structured expert elicitation using the IDEA protocol. Methods Ecol Evol. 2017 July 30;9(1):169–80.

6.  Azzolina D, Berchialla P, Gregori D, Baldi I. Prior Elicitation for Use in Clinical Trial Design and Analysis: A Literature Review. International Journal of Environmental Research and Public Health. 2021 Jan;18(4):1833.

7.  Bojke L, Soares M, Claxton K, Colson A, Fox A, Jackson C, et al. Developing a reference protocol for structured expert elicitation in health-care decision-making: a mixed-methods study. Health Technology Assessment. 2021 June 9;25(37):1–124.

8.  Gosling JP. SHELF: The Sheffield Elicitation Framework. In: Elicitation [Internet]. Springer; 2017 [cited 2024 July 22]. Available from: https://doi.org/10.1007/978-3-319-65052-4_4

9.  James A, Choy SL, Mengersen K. Elicitator: An expert elicitation tool for regression in ecology. Environmental Modelling & Software. 2010 Jan 1;25(1):129–45.

10. Johnson SR, Tomlinson GA, Hawker GA, Granton JT, Grosbein HA, Feldman BM. A valid and reliable belief elicitation method for Bayesian priors. Journal of Clinical Epidemiology. 2010 Apr 1;63(4):370–83.

11. Morris DE, Oakley JE, Crowe JA. A web-based tool for eliciting probability distributions from experts. Environmental Modelling & Software. 2014 Feb 1;52:1–4.

12. Bjornson CL, Johnson DW. Croup. Lancet. 2008 Jan 26;371(9609):329–39.

13. Brown JC. The management of croup. Br Med Bull. 2002;61:189–202.

14. Kairys SW, Olmstead EM, O'Connor GT. Steroid Treatment of Laryngotracheitis: A Meta-Analysis of the Evidence From Randomized Trials. Pediatrics. 1989 May 1;83(5):683–93.

15. Geelhoed GC, Turner J, Macdonald WB. Efficacy of a small single dose of oral dexamethasone for outpatient croup: a double blind placebo controlled clinical trial. BMJ. 1996 July 20;313(7050):140–2.

16. Yao TC, Wang JY, Chang SM, Chang YC, Tsai YF, Wu AC, et al. Association of Oral Corticosteroid Bursts With Severe Adverse Events in Children. JAMA Pediatr. 2021 July 1;175(7):723–9.

17. Bojke L, Soares MO, Claxton K, Colson A, Fox A, Jackson C, et al. Reference Case Methods for Expert Elicitation in Health Care Decision Making. Med Decis Making. 2022 Feb 1;42(2):182–93.

18. Lan J, Plint AC, Dalziel SR, Klassen TP, Offringa M, Heath A, et al. Remote, real-time expert elicitation to determine the prior probability distribution for Bayesian sample size determination in international randomised controlled trials: Bronchiolitis in Infants Placebo Versus Epinephrine and Dexamethasone (BIPED) study. Trials. 2022 Apr 11;23(1):279.

19. Clemen RT, Winkler RL. Combining Probability Distributions From Experts in Risk Analysis. Risk Anal. 1999 Apr 1;19(2):187–203.

20. European Food Safety Authority. Guidance on Expert Knowledge Elicitation in Food and Feed Safety Risk Assessment. EFSA Journal. 2014 June 19;12(6):3734.

21. Werner C, Hanea AM, Morales-Nápoles O. Eliciting Multivariate Uncertainty from Experts: Considerations and Approaches Along the Expert Judgement Process. In: Dias LC, Morton A, Quigley J, editors. Elicitation: The Science and Art of Structuring Judgement [Internet]. Cham: Springer International Publishing; 2018 [cited 2024 July 22]. p. 171–210. Available from: https://doi.org/10.1007/978-3-319-65052-4_8

22. Hiance A, Chevret S, Lévy V. A practical approach for eliciting expert prior beliefs about cancer survival in phase III randomized trial. Journal of Clinical Epidemiology. 2009 Apr 1;62(4):431-437.e2.

23. Hampson LV, Whitehead J, Eleftheriou D, Tudur-Smith C, Jones R, Jayne D, et al. Elicitation of Expert Prior Opinion: Application to the MYPAN Trial in Childhood Polyarteritis Nodosa. PLoS One. 2015 Mar 30;10(3):e0120981.

24. Thall PF, Ursino M, Baudouin V, Alberti C, Zohar S. Bayesian treatment comparison using parametric mixture priors computed from elicited histograms. Stat Methods Med Res. 2019 Feb;28(2):404–18.

25. Curran JM, Bolstad WM. Introduction to Bayesian Statistics. 3rd ed. Wiley; 2016.

26. R Core Team. R: A Language and Environment for Statistical Computing [Internet]. Vienna, Austria: R Foundation for Statistical Computing; 2024 [cited 2024 July 24]. Available from: https://www.R-project.org/

27. Chang W, Cheng J, Allaire J, Sievert C, Schloerke B, Xie Y, et al. shiny: Web Application Framework for R [Internet]. 2024. Available from: https://CRAN.R-project.org/package=shiny

28. Devleesschauwer B, Torgerson P, Charlier J, Levecke B, Praet N, Roelandt S, et al. prevalence: Tools for prevalence assessment studies [Internet]. 2022 [cited 2024 Aug 15]. Available from: https://cran.r-project.org/package=prevalence

29. Harris PA, Taylor R, Thielke R, Payne J, Gonzalez N, Conde JG. Research electronic data capture (REDCap)—A metadata-driven methodology and workflow process for providing translational research informatics support. Journal of Biomedical Informatics. 2009 Apr 1;42(2):377–81.

30.  Harris PA, Taylor R, Minor BL, Elliott V, Fernandez M, O'Neal L, et al. The REDCap consortium: Building an international community of software platform partners. Journal of Biomedical Informatics. 2019 July 1;95:103208.

31.  Klassen TP, Dalziel SR, Babl FE, Benito J, Bressan S, Chamberlain J, et al. The Pediatric Emergency Research Network (PERN): A decade of global research cooperation in paediatric emergency care. Emerg Med Australas. 2021 Oct;33(5):900–10.

32.  Hogarth RM. A note on aggregating opinions. Organizational Behavior and Human Performance. 1978 Feb 1;21(1):40–6.

33.  Aregbesola A, Tam CM, Kothari A, Le ML, Ragheb M, Klassen TP. Glucocorticoids for croup in children. Cochrane Database Syst Rev. 2023 Jan 10;(1).

34.  Parker CM, Cooper MN. Prednisolone Versus Dexamethasone for Croup: a Randomized Controlled Trial. Pediatrics. 2019 Sept;144(3):e20183772.

35.  Johnson SR, Tomlinson GA, Hawker GA, Granton JT, Feldman BM. Methods to elicit beliefs for Bayesian priors: a systematic review. Journal of Clinical Epidemiology. 2010 Apr 1;63(4):355–69.

36.  Rue H, Martino S, Chopin N. Approximate Bayesian inference for latent Gaussian models by using integrated nested Laplace approximations. Journal of the Royal Statistical Society Series B: Statistical Methodology. 2009 Apr 6;71(2):319–92.

37.  Alshehr M, Almegamsi T, Hammdi A. Efficacy of a small dose of oral dexamethasone in croup. Biomedical Research. 2005 Jan 1;16(1):65–72.

38.  Fifoot AA, Ting JY. Comparison between single-dose oral prednisolone and oral dexamethasone in the treatment of croup: A randomized, double-blinded clinical trial. Emergency Medicine Australasia. 2007;19(1):51–8.

39.  Higgins JPT, Thompson SG, Spiegelhalter DJ. A re-evaluation of random-effects meta-analysis. Journal of the Royal Statistical Society: Series A (Statistics in Society). 2009;172(1):137–59.

40.  Jackson D, Bowden J, Baker R. How does the DerSimonian and Laird procedure for random effects meta-analysis compare with its more efficient but harder to compute counterparts? Journal of Statistical Planning and Inference. 2010 Apr 1;140(4):961–70.

41.  Janis IL. Victims of Groupthink: A Psychological Study of Foreign Policy Decisions and Fiascoes. Houghton Mifflin Co; 1972.

**Clinical Case Study for Elicitation**

Consider a previously healthy 1-year-old boy who presents to the ED in the autumn. He has been previously well and developed a runny nose a few days ago along with a mild fever. Today he developed noisy breathing at home with a barking cough. His oral intake has slightly decreased but he is voiding well. You notice mild but audible stridor with a barking cough when you see him in the emergency department. He has mild intercostal and subcostal indrawing. His oxygen saturation is 97%, his heart rate is 140 beats per minute, and his respiratory rate is 25/minute. You make a diagnosis of mild to moderate croup. You decide to treat him with dexamethasone.

Question 1: Given the clinical guidelines you opt to provide dexamethasone at a dose of 0.60 mg/kg. Out of 100 similar patients with mild croup, how many will revisit the ED within 7 days following this visit if treated with this dose?

Question 2: Given the clinical guidelines and some pilot work, you opt to provide dexamethasone at a dose of 0.15 mg/kg. Out of 100 similar patients with mild croup, how many will revisit the ED within 7 days following this visit if treated with this dose?

**Physician's Professional Experience Survey**

How many years have you practiced medicine?

How many years of experience in pediatric emergency medicine?

Have you prescribed 0.6 mg/kg of dexamethasone for croup over the last year (yes/no)?

Have you prescribed 0.15 mg/kg of dexamethasone for croup over the last year (yes/no)?

What's the maximum dose (mg) of dexamethasone you have prescribed for croup in the last year?

Have you received any training in clinical trial methodology (yes/no)?

Have you received any training in statistics (yes/no)?

**Elicitation Workshop Script**

**Introduction**
Hi everyone, welcome to the elicitation workshop for the CROUP Dosing Trial.
My name is Arlene Jiang – I will be your workshop facilitator alongside Apoorva for today. This workshop will be conducted a total of three times, in order to elicit beliefs from experts across different regions and time zones.

**Page 2**
The CROUP Dosing Trial is a double-blind, RCT comparing two different doses of dexamethasone for children presenting to the emergency department with croup. It aims to determine whether a lower dose of dexamethasone is non-inferior to the standard dose. We will investigate how these treatment regimens impact the number of return visits to the ED or readmissions to the hospital within 7 days following initial presentation to the emergency department. The purpose of this workshop is to formulate a prior probability distribution that can be used to calculate the sample size of the CROUP Dosing Trial using Bayesian methods.

**Page 3**
In place of standard frequentist statistics, this trial will implement Bayesian methods. Frequentist sample size determination require assumptions for significance (think less than 5% error), power (80% is typically deemed acceptable), and an effect size of interest. These assumptions are complex and prone to error given that there is a need for a trial and limited knowledge in these areas.
Bayesian statistics incorporate prior information into analyses. This prior information can come in the form of data or expert opinion. In this case, prior information will be derived from your expert opinion and experience in pediatric emergency medicine and croup. We will aggregate your individual distribution to form a prior distribution, to represent all of your opinions. These aggregated expert opinions will form prior probabilities that will be used to determine the sample size of the CROUP Dosing trial, using Bayesian methods.

**Unshare Screen**
Today we have 6 (3 for workshop 2 and 3) experts. Could we start with a round of introductions? Thanks so much for your time - we really appreciate your participation.

**Share Screen – page 5**
The workshop will consist of two rounds of surveys, separated by a short break and a group discussion. Each survey round will include 2 questions to elicit your beliefs. We will also collect information about your professional experience through a short questionnaire.

**Page Consent**
Today's workshop involves minimal risk. As a benefit, you will gain more knowledge in the treatment of croup and will be provided with study results to see the quantitative expression of your subjective judgement. We hope that the information learned from this study can be used in the future to benefit other researchers who are aiming to understand how croup treatments are viewed by experts. The output from this study will also be used to support the design and analysis of the CROUP Dosing study, which will generate evidence regarding the treatment of children with croup.

If you have any questions about this project, you may contact me or Apoorva or Alex or Terry, we are very happy to provide any additional information. By completing these questionnaires, you are consenting to its use in research.

All information collected about you will be "de-identified" by replacing your identifiable information (i.e., name) with a "study number". Only the "study code key" can connect the information collected about you to your identity. The study code key will be safeguarded by the research team at Children's Hospital Research Institute of Manitoba. Even though the risk of identifying you from the study data is very small, it can never be completely eliminated.

## Page 6

Prior to today's workshop, you should have received an email containing the following materials for today's workshop. First, you should now have a unique user ID. You will need to submit this ID alongside all of your answers in today's workshop. You should have also received a link to the elicitation tool and our professional experience questionnaire. The elicitation tool can only be accessed using a computer or laptop.

## Page 7

Our workshop will primarily be hosted on the Elicitation Tool on R-Shiny Dashboard. The menu on the left-hand side will guide you through each survey round.

## Page 8

Each round will consist of 4 steps, 2 questions and 2 submission pages.

## Page 9

Here is a sample question page.

## Page 10

1. Read the instructions
2. Read the question.
3. Use the slider to choose the lower, upper and best plausible value
4. Check the summary and plot
5. If the summary and plot match your opinion, enter the values in Step 2.

I would recommend writing down the selected values.

## Page 11

1. Select the next Step 2 – Enter your values – Question 1 tab on the menu bar to enter your answers.
2. Enter the Unique User ID emailed to you.
3. Enter the Lower, Upper and Best Plausible value based on your opinion in Step 1.
4. Submit the values.
5. Follow the same steps in STEP 3 AND STEP 4

Be sure to hit submit. You should see a pop-up confirmation box.

Demo - https://apoorvagangwani.shinyapps.io/EXAMPLE/
Now let's go through an example. After reading the instructions, let's review the question. You're a tech product manager, and you're deciding between two email notification frequencies for a mobile app.

Given the standard notification frequency of once a day. Out of 100 app users who receive daily email notifications, how many will unsubscribe within a week due to email overload?

Step 1, here, is to choose the lower and upper plausible values for the number of people who unsubscribe.

Based on my experience, I tend to hear of people receiving too many emails, incentivizing unsubscribing. On the other hand, people may not unsubscribe for a variety of reasons: maybe they have email filters, or do not check their email frequently. Furthermore, I will consider that we would be interested in people unsubscribing within a week for this question. Based on these factors, I will select 1 as the lower plausible value and 40 as the upper plausible value. Next, I'll select the most likely value. Notice how adjusting the slider changes which answers are most probable. I think a most likely value of 7 best represents my beliefs. Then, I'll record these values: 1, 40, 7 to enter on the next page. I will enter each of these values on this page alongside my user ID and hit submit – prompting the pop-up message.

And so on for question 2.

## Page 12

The formal elicitation process will be presented in a similar format. Once everyone has completed and submitted their answers for both questions, there will be a short break. If you have not already, please use this time to fill out the professional experience survey and take some time away from your screen. Apoorva will generate a deidentified boxplot with everyone's answers that will be discussed for 10-15 minutes. Please share your thoughts with other group members. Remember that it is not necessary to reach a consensus across the group, instead use the boxplot and discussion to calibrate your judgements. In the second questionnaire, you may adjust your answers.

## Page 13

We will begin the formal elicitation now. Please open the Elicitation Tool link in your email. If you haven't received it, it may be in your junk box or just let me know. Is everyone able to open step 1 of round 1? Ok, please get started with the round of questions. You will have 8 minutes to complete the round. Be sure to accurately enter the values on the submission pages. If you finish early, please use this time to fill out the professional experience questionnaire. Message a 1 if you have completed the survey or if you have any questions.

## Page 14

We're now moving into the discussion stage. You should have completed the professional experience questionnaire at this time as well. During this stage, we will be presenting a boxplot for each answer. Each boxplot will display a median (50% of patients are under this value and 50% over). The two lines represent whiskers. Longer whiskers represent more conservative opinions.

Discuss for 15 minutes
- What information did you draw on to select your central tendency?
- What decision-making process led you to choose your estimates?

- <span style="color:red">What factors may have introduced different levels of uncertainty in the distributions?</span>
- <span style="color:red">If you are willing to share which distribution you produced, why do you think yours differs from the others presented?</span>

**Page 15**

Let's move on to our last step here the second-round questionnaire. Please navigate back to the Elicitation Tool. If the page has disconnected, please refresh the page.

This second-round will present the same questions as round 1. However, these answers will be used in our final analyses. You may change your answers from round 1. Be sure to accurately enter the values on the submission pages and hit submit. Please let me know if you have any questions, and you may start the second round now. Type a 2 in the chat once you have completed your survey.

**Close**

It looks like everyone has completed the final questionnaire. Since our last workshop is on April 16[th] at 12pm Eastern Time, we will circulate an individual and group distribution two weeks after that workshop. Please also let our team know if you do not wish to be acknowledged in the final publication. If you have any questions, please feel free to email me or Apoorva or Alex.

Thanks so much for your participation and discussion today.

**Deidentified boxplots of expert opinion**

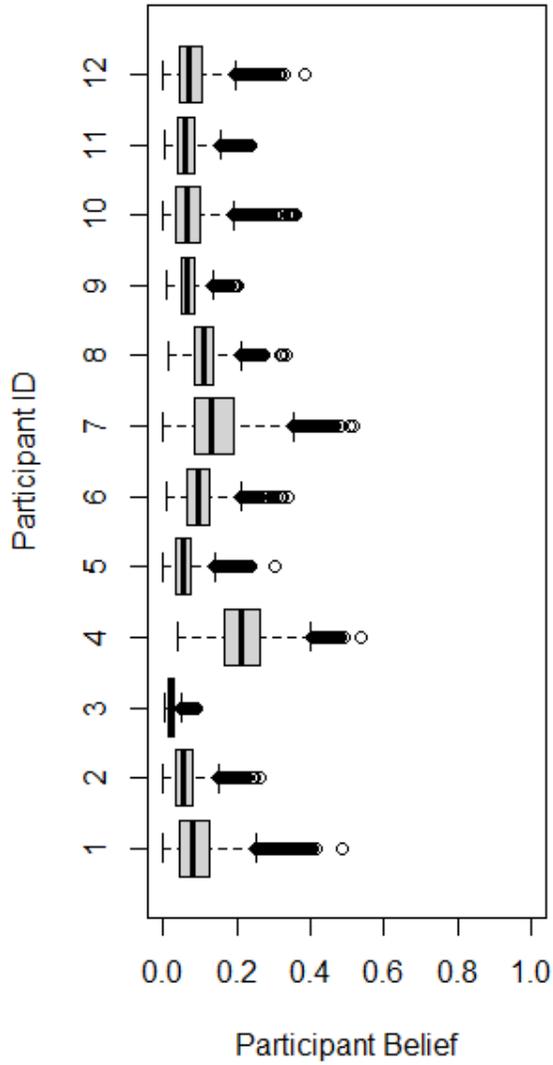

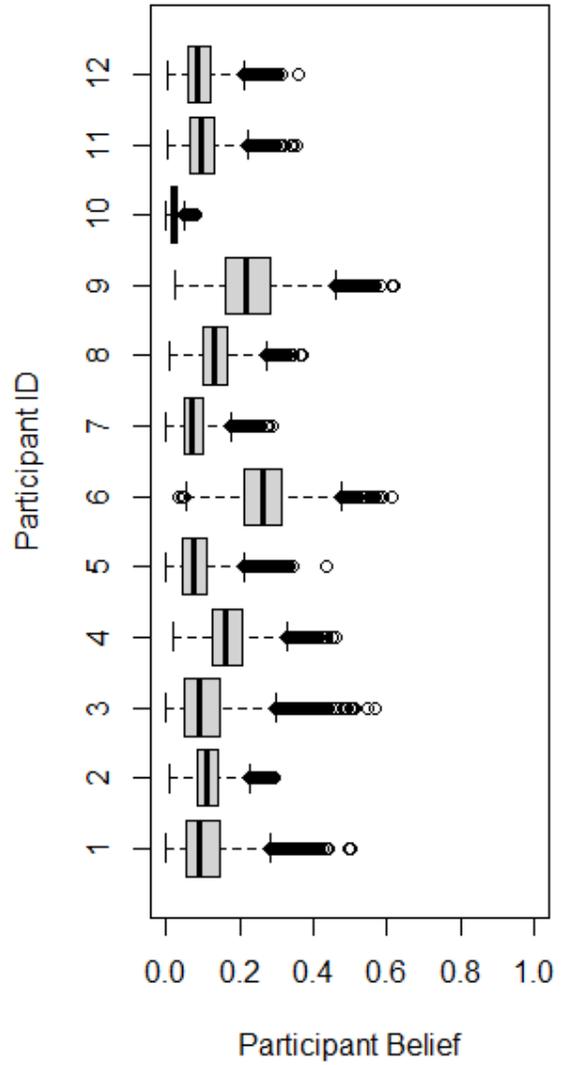

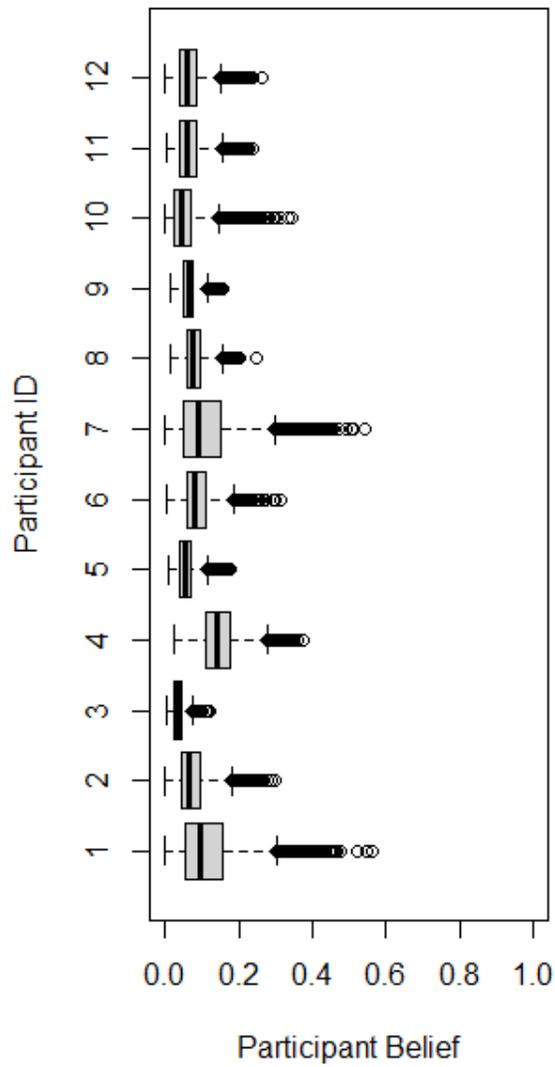

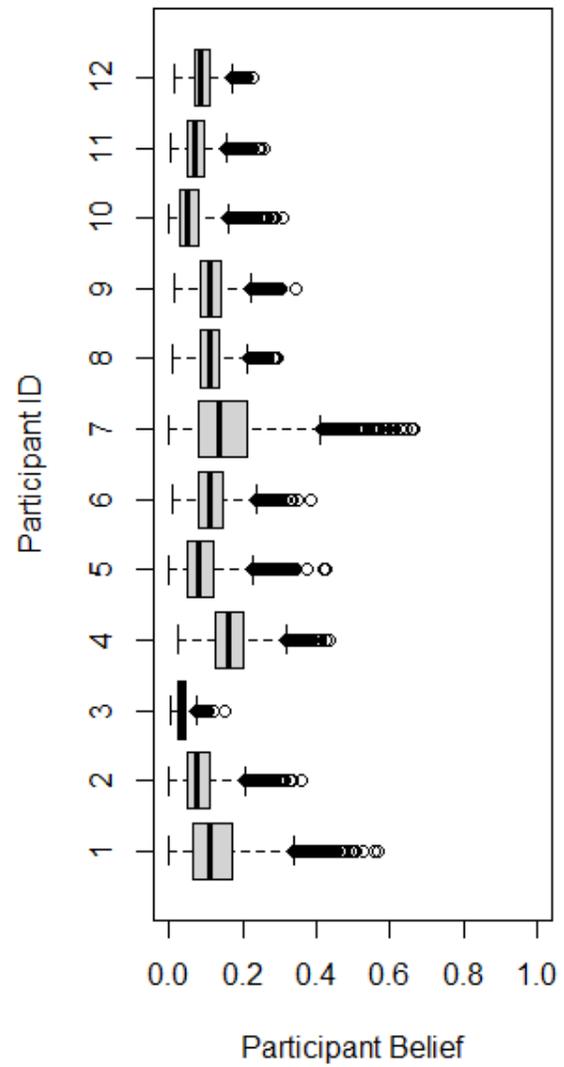

**Physician's survey to determine non-inferiority margin of using dexamethasone at**

**0.15mg/kg to treat croup**

A case scenario

Except for most severe children who cannot tolerate orally, consider 100 children aged 6 months to five years with croup who present in the Emergency Department (ED). You believe that they should be treated with dexamethasone. You have the option of providing dexamethasone at either 0.6mg/kg or 0.15mg/kg. Under the standard dose of 0.6 mg/kg, 8 children, on average, return to the emergency department. Because pilot work suggests that 0.15 mg/kg is similarly effective with fewer adverse events (i.e. bronchopneumonia, sepsis, bacteria tracheitis, febrile convulsion, GI bleeding distress and disseminated varicella infection), you opt to prescribe 0.15 mg/kg of dexamethasone. Keeping in mind the reduced side effect profile of 0.15 mg/kg and that you can repeat dexamethasone prescription if the treatment fails, how many additional patients would you accept returning to the ED to routinely prescribe 0.15 mg/kg compared to 0.6 mg/kg?

Non-inferiority margin:

| Non-inferiority Margin | Number of Responses | Non-inferiority Margin | Number of Responses |
|---|---|---|---|
| 1 | 6 | 9 | 0 |
| 2 | 20 | 10 | 6 |
| 3 | 3 | 11 | 0 |
| 4 | 2 | 12 | 3 |
| 5 | 10 | 13 | 0 |
| 6 | 2 | 14 | 0 |
| 7 | 0 | 15 | 2 |
| 8 | 6 | | |